\documentclass[
	pra,
	aps,
	reprint,
	amsmath,amssymb,
	a4paper,
	superscriptaddress,
	10pt,
    nofootinbib
]{revtex4-2}

\usepackage[english]{babel}
\usepackage[utf8x]{inputenc}
\usepackage[T1]{fontenc}
\usepackage{notoccite}
\usepackage{gensymb}
\usepackage{siunitx}
\usepackage{caption}
\usepackage{caption}
\usepackage{float}
\usepackage{float}
\usepackage{amsmath}
\usepackage{graphicx}
\usepackage[colorinlistoftodos]{todonotes}
\usepackage[colorlinks=true, allcolors=blue]{hyperref}
\usepackage{subfiles}
\usepackage{subcaption}
\usepackage{tabularx}
\usepackage{graphicx}
\usepackage{graphicx}
\usepackage{tikz}

\begin{document}

\title{Deterministic single-photon sources in hexagonal boron nitride with electron-dose-tuned purity and reversible thermal quenching}

\author{Amrita Majumder}
\affiliation{%
Laboratory of Optics of Quantum Materials, Department of Physics, Indian Institute of Technology Bombay, Mumbai- 400076, India 
}

\author{Janhavi Khunte}
\affiliation{%
Laboratory of Optics of Quantum Materials, Department of Physics, Indian Institute of Technology Bombay, Mumbai- 400076, India 
}

\author{Ikshvaku Shyam}
\affiliation{%
Laboratory of Optics of Quantum Materials, Department of Physics, Indian Institute of Technology Bombay, Mumbai- 400076, India 
}

\author{Rohit Kumar}
\affiliation{%
Laboratory of Optics of Quantum Materials, Department of Physics, Indian Institute of Technology Bombay, Mumbai- 400076, India 
}

\author{Anshuman Kumar}
\email{anshuman.kumar@iitb.ac.in}
\affiliation{%
Laboratory of Optics of Quantum Materials, Department of Physics, Indian Institute of Technology Bombay, Mumbai- 400076, India 
}%
\affiliation{Centre of Excellence in Quantum Information, Computation, Science and Technology, Indian Institute of Technology Bombay, Powai, Mumbai 400076, India}

\begin{abstract}
Electron-beam irradiation is an established route to create site-controlled,
room-temperature single-photon emitters (SPEs) in hexagonal boron nitride
(hBN), but two aspects remain underexplored: how the electron dose governs the
properties of the resulting \emph{single} emitters, and how the emission
behaves when the host is heated above room temperature. Here, we create emitters
deterministically with a focused electron beam and confirm single-photon
emission across three independent flakes, with $g^{(2)}(0)=0.09$, $0.12$, and
$0.16$. We map the single-emitter response (yield, spectrum, lifetime, and photon
purity) as a function of electron dose, identifying an optimal
window for high-purity single emitters. Consistent with recent cryogenic
studies, we assign the bright room-temperature feature near \SI{575}{nm} to the
phonon sideband (PSB) of a green--yellow emitter whose zero-phonon line (ZPL)
lies near \SI{548}{nm}. Temperature-dependent photoluminescence measured \emph{in situ} under real-time from room
temperature to \SI{300}{\degreeCelsius} reveals a thermal quenching that is
fully reversible upon cooling, in contrast to the irreversible
annealing-induced degradation reported elsewhere, indicating that transient
heating does not permanently damage the centers. These results add quantitative dose control and above-room-temperature operation to
the toolbox for deterministic hBN quantum-light sources.


\end{abstract}

\maketitle
\section{Introduction}
The field of quantum photonics is rapidly expanding, driven by its potential to transform existing information and communication technology\cite{Shields2007}. It is concerned with the controlled generation and use of light at the level of individual photons, which serve as fundamental bearers of quantum information\cite{Gisin2007, Kimble2008}. Such exact control over light enables important applications such as secure data transmission, quantum information processing, and high-sensitivity measurements. A key requirement for these technologies is the availability of stable and efficient single photon sources that can operate under real-world conditions, notably at room temperature, while maintaining high emission rates and interoperability with integrated photonic systems\cite{Singh2026,Zalogina2025,Aharonovich2016}. Solid-state quantum emitters implanted in wide bandgap materials have emerged as promising possibilities due to their durability, scalability, and compatibility for device integration\cite{Aharonovich2014, Aharonovich2016, awschalom2018quantum}.

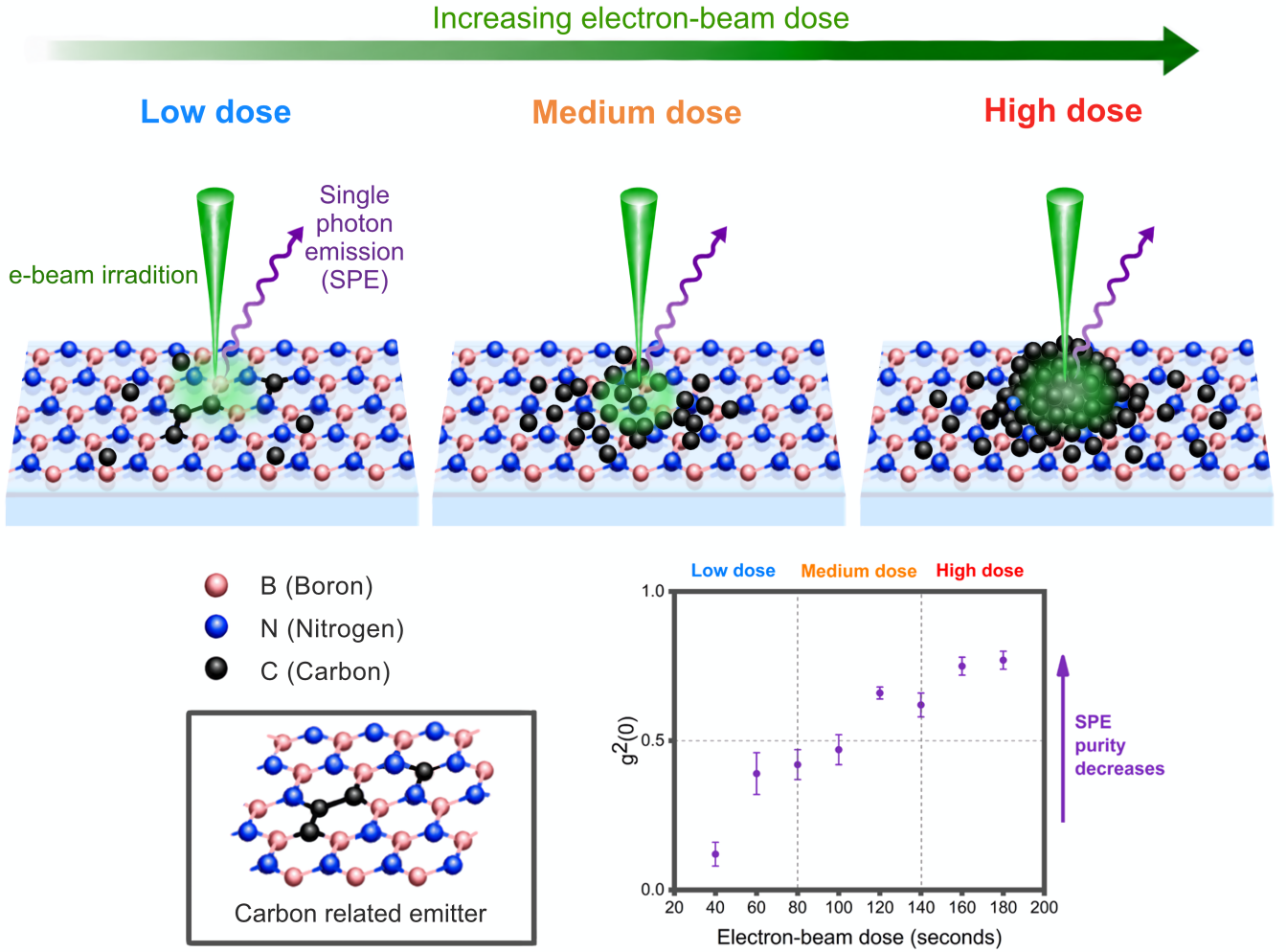
\begin{figure*}[t]
\centering

\begin{picture}(0,0)
\put(-5,80){\textbf{}}
\end{picture}
\includegraphics[width=1\textwidth]{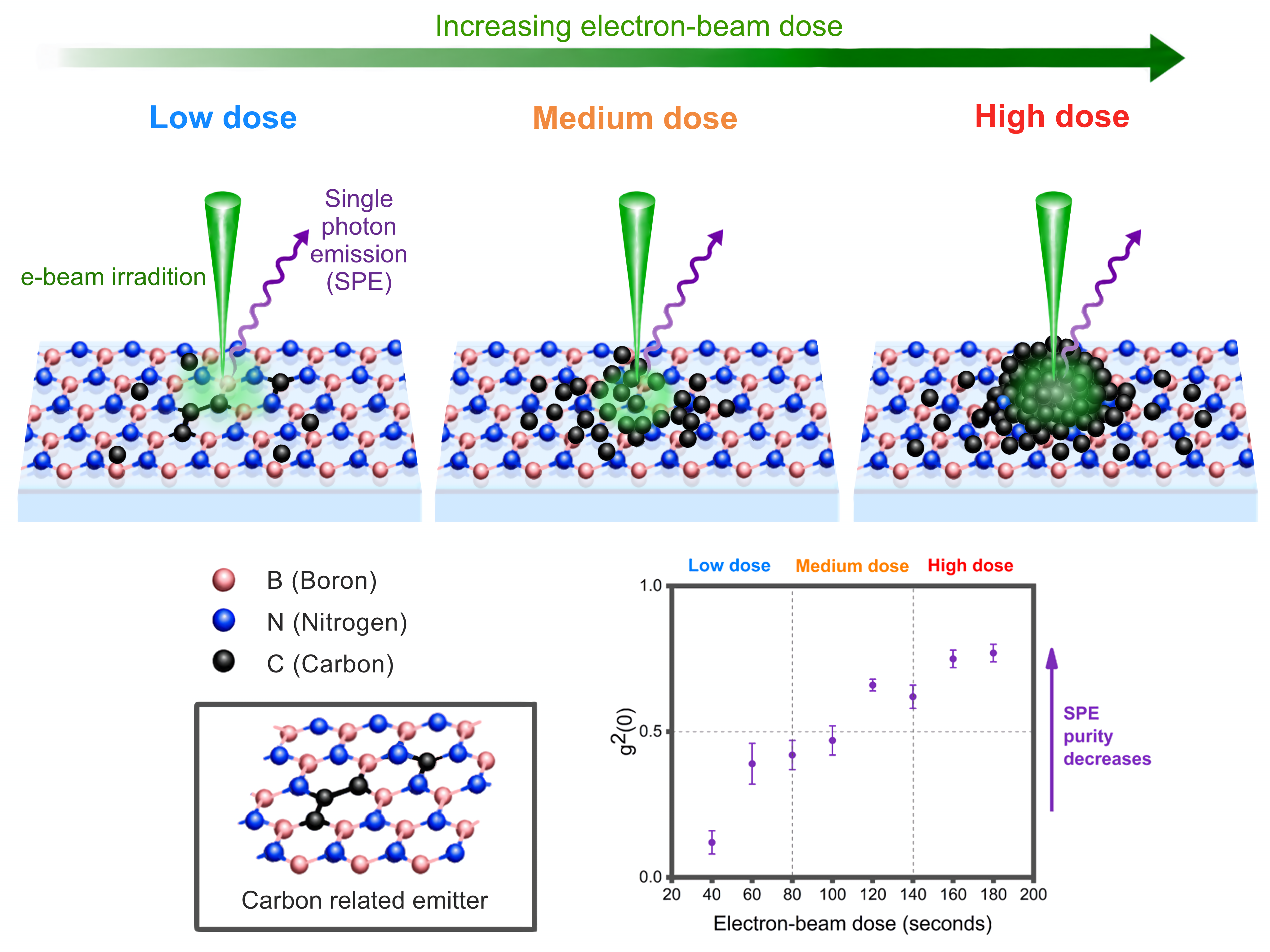}\hspace{0.015\textwidth}

\caption{{A schematic representation of electron-beam-irradiated hBN, demonstrating the progression of carbon deposition and the change in single-photon emitter purity as the irradiation dosage increases.}
}
\label{fig:figure1}
\end{figure*}

Hexagonal boron nitride (hBN) hosts optically active point defects that serve as bright, photostable single-photon emitters (SPEs) capable of operating at room temperature\cite{tran2016quantum, exarhos2017optical, jungwirth2016temperature}. Because these SPEs reside within an atomically thin van der Waals material, they are highly promising components for integrated quantum photonics. To utilize them in practical technologies, a reliable method for deterministic creation—positioning an emitter at a precise location with predictable characteristics—is essential. Recent studies have demonstrated substantial progress in the development of quantum emitters based on hBN. These advances include room-temperature single-photon emission, tunability of emission characteristics through strain engineering, and successful integration with photonic components such as waveguides and optical resonators\cite{grosso2017tunable, toth2019single, dietrich2018observation, singh2025plasmonic, sakib2024purcell}. Various approaches, including focused ion beam techniques and electron irradiation, have been explored to intentionally introduce defects in hBN, highlighting the potential for controlled emitter fabrication\cite{fournier2021position, shotan2016photoinduced, chejanovsky2016structural, kumar2023localized,arxiv.2606.12304}. Among these methods, electron-beam irradiation within a scanning electron microscope (SEM) has emerged as a widely adopted, mask-free technique for this purpose. Specifically, treating hBN with this type of irradiation reliably generates a green-yellow emitter, a process whose yield and spatial precision have been thoroughly established \cite{proscia2018near, kumar2023localized}.

To fully realise these emitters' potential as tailored optical sources, two critical research gaps must be filled. First, while the successful generation and spatial localisation of these defects have been widely documented\cite{tran2016quantum, proscia2018near, chejanovsky2016structural, kumar2023localized}, there is a paucity of systematic data on how the applied electron dose(irradiation time) affects critical single-emitter performance metrics. The link between electron dose and properties such as formation probability, emission intensity, single-photon purity $g^{(2)}(0)$, and spectral profile (position and linewidth) is not fully mapped. Previous studies of electron dosage have been focused on macroscopic ensemble averages or spin-defect concentrations\cite{ shotan2016photoinduced, ngoc2018effects}. Second, current thermal assessments of these emitters are incomplete. Previous research has mostly examined them under cryogenic circumstances, allowing researchers to detect electron-phonon interactions and resolve the zero-phonon line (ZPL)\cite{hazra2026temperature}, or through ex-situ annealing to determine permanent heat resistance\cite{hazra2026insights}. Crucially, the real-time, in-situ dynamics of these emitters when exposed to temperatures higher than room temperature remain completely undocumented.

A recent spectral reassignment clarifies the identity of the dominant room‑temperature feature near 575 nm. Cryogenic and hyperspectral measurements have shown that this peak, historically labeled the zero‑phonon line (ZPL), is in fact the phonon sideband (PSB); the true ZPL lies near 548 nm and is strongly broadened at ambient temperature, so the PSB dominates the spectrum\cite{hazra2026temperature, hazra2026insights}. We adopt this corrected assignment throughout. Establishing a precise dose threshold for the regulated creation of quantum emitters becomes crucial since electron irradiation might cause the deposition of a carbonaceous layer and affect the emitter purity. Specifically, in this study,  we (i) map the single‑emitter response as a function of electron dose, (ii) demonstrate the reversibility of thermal quenching up to $300\degree C$, establishing the structural resilience of the emitters against the permanent emission loss and irreversible degradation induced by high-temperature ex-situ annealing, (iii) demonstrate reproducible single‑photon emission from electron‑beam‑created emitters across multiple hBN flakes, and (iv) characterize the photophysics of an emitter using the corrected ZPL/PSB assignment.

\begin{figure*}[t]
\centering

\begin{picture}(0,0)
\put(-5,80){\textbf{}}
\end{picture}
\includegraphics[width=0.93\textwidth]{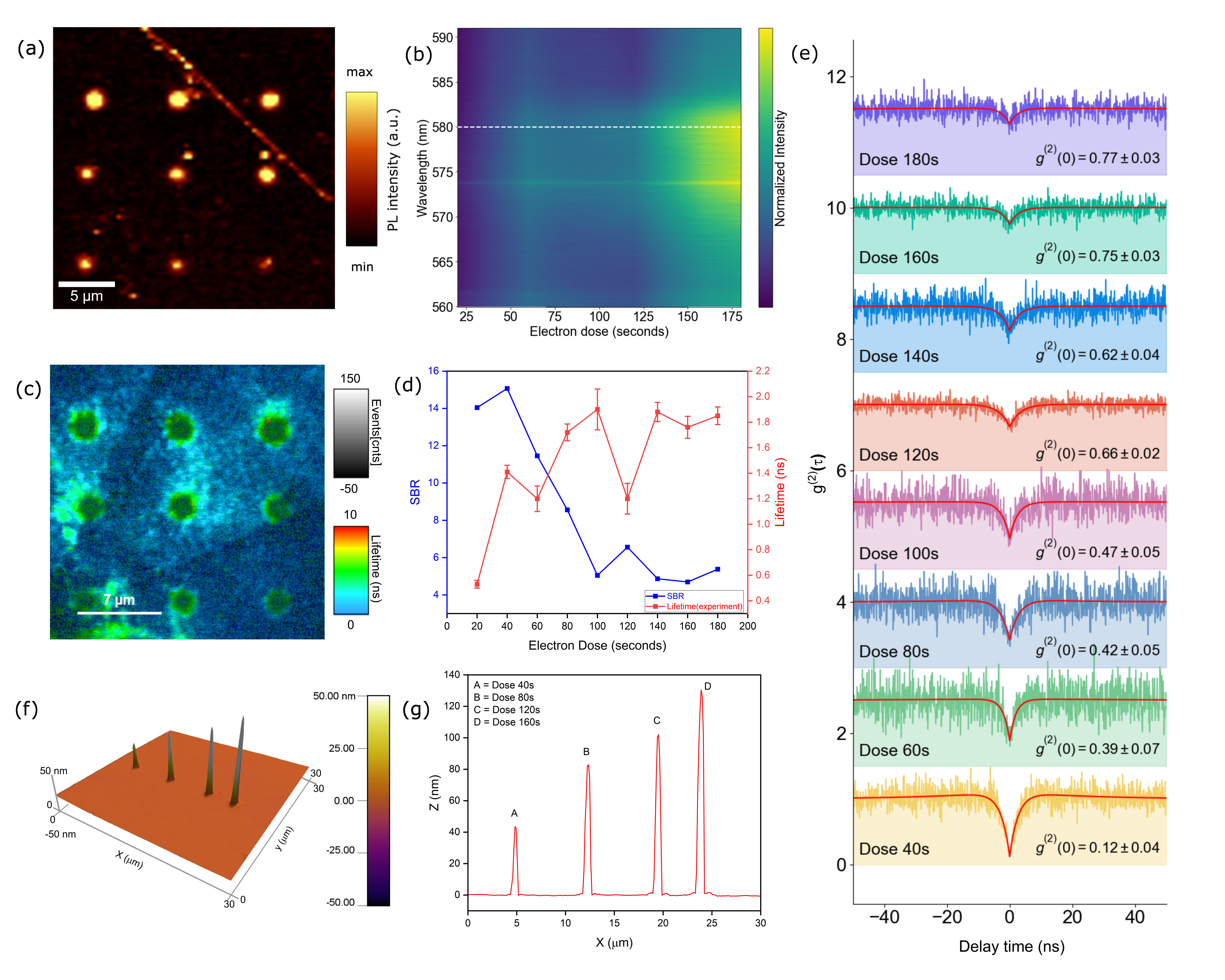}\hspace{0.015\textwidth}
\caption{\textbf{Dose-dependent effects of electron-beam irradiated color centers in hBN.}
\textbf{a}, Confocal photoluminescence (PL) map of e-beam irradiated hBN flake, where the irradiation duration was varied from 20s to 180s. The map was acquired under 532 nm laser excitation.
\textbf{b}, Color-coded 2D map showing the PL intensity as a function of emission wavelength and electron beam irradiation time (20–180 s) under 532 nm excitation. A broad emission band centered near $\sim 575$ nm, attributed to the phonon sideband (PSB) of the emitter, is observed across all doses. A narrow feature at $\sim 574$ nm corresponds to the intrinsic hBN $E_{2g}$ phonon mode (Raman shift $\sim 1366\ \mathrm{cm}^{-1}$). With increasing irradiation time, a broad band centered near $\sim 580$ nm (white dashed line) progressively emerges and intensifies, consistent with the graphitic Raman G-band, indicating increased graphitization and formation of $\mathrm{sp}^2$-bonded carbon\cite{ferrari2006raman, dresselhaus2010perspectives}. 
\textbf{c}, Lifetime-weighted fluorescence image of hBN emitters obtained under pulsed 532 nm laser excitation with a repetition rate of 20MHz.
\textbf{d}, The electron dose-dependent extracted excited-state lifetime and the signal-to-background ratio (SBR), defined as the ratio between the integrated intensity of the emitter's signal ($\sim$ 575nm) and the carbon-related emission ($\sim$ 580nm).
\textbf{e}, Dose-dependent second-order autocorrelation measurements,$g^{(2)}(\tau)$, yielding lower emitter's purity with increasing irradiation dose.
\textbf{f-g}, AFM topography of irradiated regions, showing the formation of nanoscale surface features whose dimensions increase with irradiation dose.
}
\label{fig:figure2}
\end{figure*}
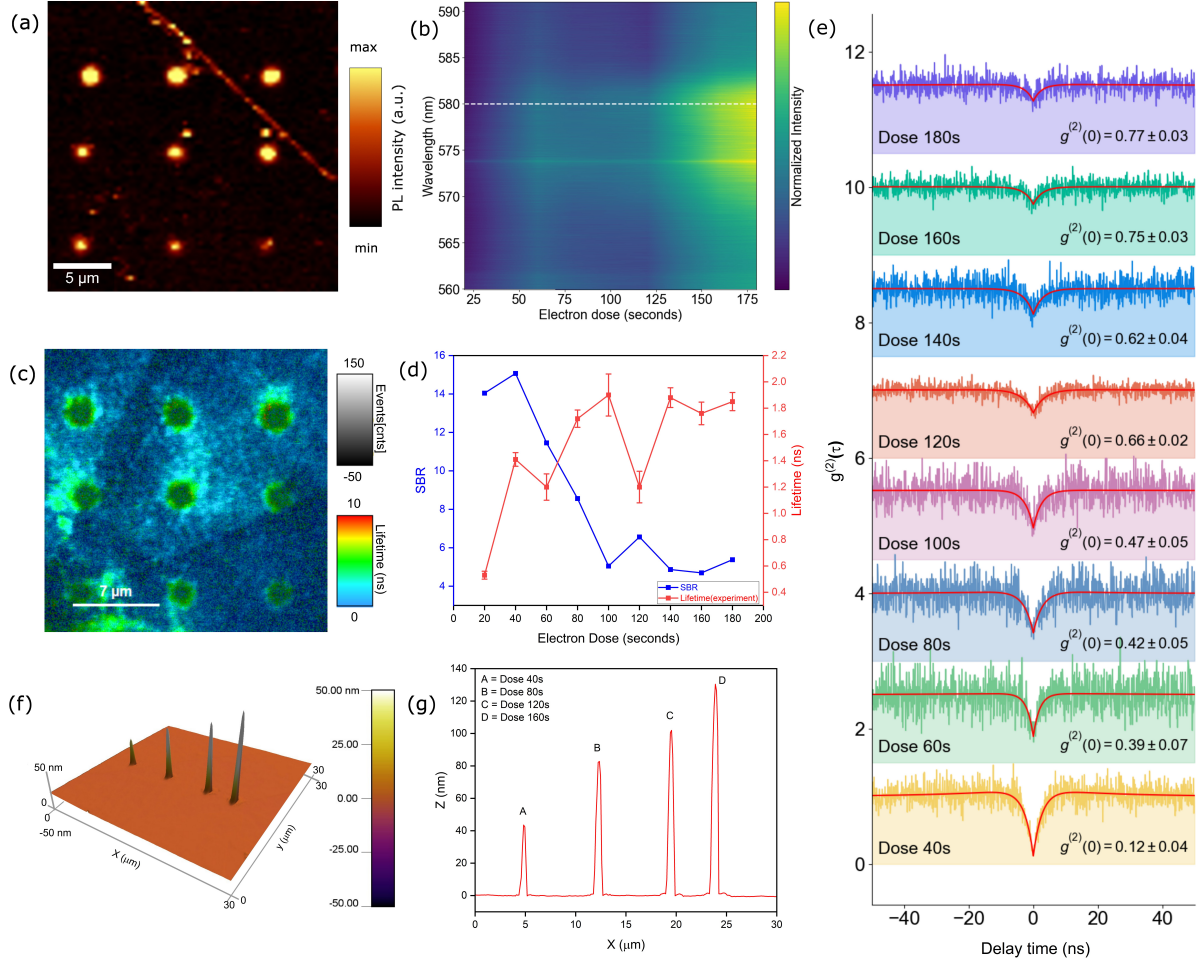

\section{Results and discussion}
Hexagonal boron nitride (hBN) flakes were mechanically exfoliated from bulk crystals (HQ Graphene) using Nitto blue tape and a PDMS stamp. These flakes were subsequently placed onto a grid-patterned Si/SiO$_2$ substrate (featuring a 300 nm thermal oxide layer) via a dry-transfer process (see Methods). To create site-specific quantum emitters, we performed localized electron-beam irradiation using scanning electron microscopy (SEM). This technique leverages moderate-energy electrons to induce localized lattice defects that act as stable color centers. Fig. 1 shows a schematic illustration of electron-beam–irradiated hBN showing the evolution of carbon deposition and the corresponding single-photon emitter purity with increasing irradiation dose. To avoid creating additional, randomly located emitters during imaging, we did not perform SEM imaging of the flakes that underwent optical characterization. 

Fig. 2 presents a systematic investigation of the influence of electron-beam irradiation dose (controlled via exposure time) on the formation, optical response, and structural environment of single-photon emitters (SPEs) in hexagonal boron nitride (hBN). We employed a confocal photoluminescence (PL) microscopy setup to investigate the optical characteristics of the irradiated arrays. All measurements were conducted at room temperature using a 532 nm continuous-wave (CW) laser with an excitation power of 100 µW, unless stated otherwise. To suppress the excitation signal, a 550 nm long-pass filter was introduced in the collection path. The confocal PL map in Fig. 2a shows clearly defined bright emission spots confined to the irradiated regions(electron dose varied from 20s to 180s); notably, these emitters are produced under conditions distinct from those reported in earlier studies\cite{kumar2023localized, tran2016robust}. These measurements correspond to defect centers generated under irradiation conditions of 48.2 pA beam current, 10 kV accelerating voltage, and (20s to 180s) electron dose. In Fig. 2b, a color-coded 2D map showing photoluminescence (PL) spectra acquired for electron doses ranging from 20s to 180s reveals a clear dose-dependent evolution. Alongside the emission from the color center quantum emitter near $\sim$575 nm with increasing exposure, a pronounced photoluminescence feature near $\sim$580 nm (corresponding Raman shift near $1580$ cm$^{-1}$) progressively emerges and intensifies, which is characteristic of graphitic ($sp^2$-bonded) carbon\cite{ferrari2006raman, dresselhaus2010perspectives}. This trend indicates the gradual accumulation of a carbonaceous layer on the irradiated regions. To quantify this effect, Fig. 2(d) plots the signal-to-background ratio (SBR), defined as the ratio between the integrated intensity of the emitter's signal ($\sim$ 575 nm) and the deposited carbon-related emission ($\sim$ 580 nm). The SBR initially remains relatively high at lower doses, indicating the dominant emitter's signal. However, with increasing irradiation time, the SBR decreases significantly, reflecting the increasing contribution of carbon-related luminescence.

Time-resolved fluorescence measurements provide further insight into the emitter dynamics. The lifetime weighted fluorescence image of hBN emitters is shown in Figure 2c, while Fig. 2(d) summarizes the dependence of the extracted excited-state lifetime on electron dose. An overall increasing trend in lifetime is observed with dose, suggesting two possibilities: one is the passivation of non-radiative decay channels, where the measured fluorescence lifetime $\tau$ is determined by the sum of the radiative decay rate ($\Gamma_{rad}$) and the non-radiative rate ($\Gamma_{non-rad}$), expressed as:
\begin{equation}
    \tau = \frac{1}{\Gamma_{\text{rad}} + \Gamma_{\text{non-rad}}}
\end{equation}
Bare structural vacancies or surface damage frequently have "dangling bonds" or "local charge traps"\cite{tran2016quantum} that give quick, non-radiative paths for excited electrons to relax, keeping $\tau$ relatively tiny. Carbon deposits and integrates into the defect site, which can passivate these trap states. By decreasing $\Gamma_{non-rad}$, the observed lifespan naturally increases and becomes dominated by the carbon center's radiative transition. And second is the alteration of the local dielectric environment. The physical deposition of carbon on the hBN sample alters the local dielectric environment; this is most likely a secondary consequence in comparison to the defect transformation. A dipole's radiative emission rate is affected by the refractive index of the surrounding medium (related to the Purcell effect)\cite{vogl2019atomic}. The deposition of an amorphous carbon layer changes the local density of optical states (LDOS) near the color center defect, which can affect the lifespan, though this usually causes smaller shifts.

The quantum nature of the emitters is evaluated at room temperature using a Hanbury-Brown and Twiss (HBT) interferometer\cite{brown1956correlation}. The second-order correlation measurements, presented in Fig. 2(e), were performed on these irradiated sites at an excitation power of $ 100~\mu\text{W} $. These values, obtained without any background subtraction or dark-count correction, were fitted using a three-level system model(\cite{singh2025plasmonic, kitson1998intensity},
\begin{equation}
  g^{(2)}(\tau) = 1 - ae^{-\tau/\tau_e} + be^{-\tau/\tau_m}
\end{equation}
,where $\tau _e$ and $\tau_m$ are excited and metastable state lifetimes, respectively.
The SPEs exhibiting antibunching behavior (second-order correlation $g
^{(2)}(0) < 0.5$) are consistently observed for doses up to $\sim100 s$, confirming the preservation of single-photon purity within this regime. At higher doses, however, the increasing carbon background introduces additional luminescence contributions that degrade emitter purity, as corroborated by the second-order correlation measurements.

Atomic force microscopy (AFM) measurements in Fig. 2(f-g) further quantify the topographical consequences of irradiation. Height profiles corresponding to doses of 40s, 80s, 120s, and 160s clearly show the formation of nanoscale protrusions at the irradiated sites with peak heights of 43.47 nm, 82.82 nm, 102.14 nm, and 130.94 nm, respectively. Both the number and height of these features increasing with dose, indicating progressive surface modification under electron irradiation. These protrusions are consistent with localized carbonaceous buildup induced by prolonged electron exposure\cite{utke2008gas, van2008critical}. The quantitative height analysis confirms a monotonic increase in feature size with irradiation time, reinforcing the interpretation of dose-driven deposition. Moderate electron doses $(\leq 100 s)$ enable the deterministic formation of high-purity single-photon emitters, whereas excessive exposure leads to significant carbon deposition that degrades emitter quality.

\begin{figure*}[t]
\centering

\begin{picture}(0,0)
\put(-5,80){\textbf{}}
\end{picture}
\includegraphics[width=0.95\textwidth]{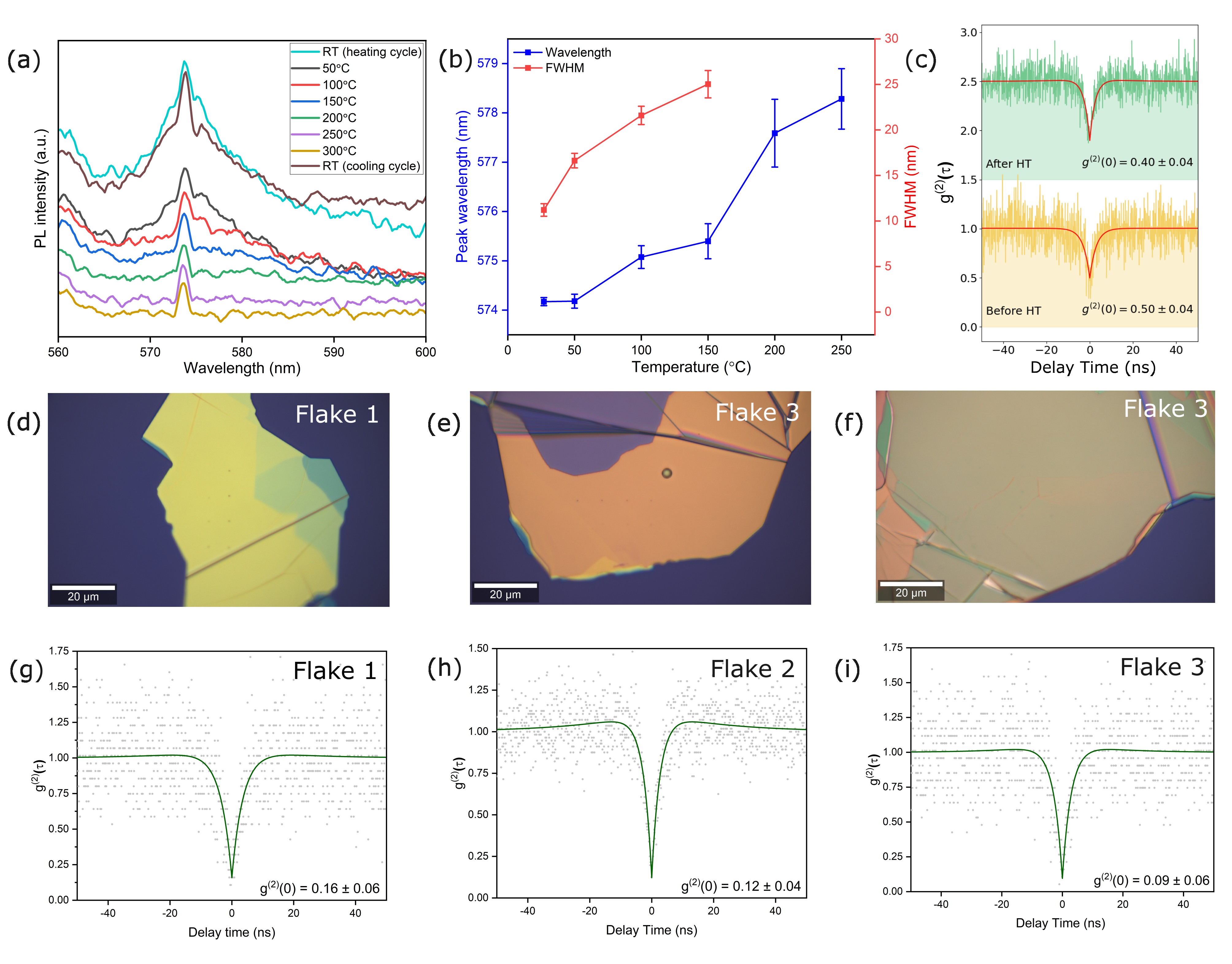}\hspace{0.015\textwidth}

\caption{\textbf{Reversible thermal response of single-photon emitters and second-order autocorrelation measurements,$g^{(2)}(\tau )$ of hBN flakes of different thickness values.}
\textbf{a}, Photoluminescence spectra were measured during heating from room temperature to $300^\circ\mathrm{C}$ and after cooling back to room temperature. The emission exhibits progressive thermal quenching with increasing temperature while retaining its spectral signature, and recovers upon cooling, indicating reversible thermal behaviour.
\textbf{b}, Temperature dependence of the emission wavelength and FWHM of the emitter.
\textbf{c}, Room-temperature second-order autocorrelation functions measured (on emitter of irradiation dose 100s) before and after thermal cycling. Fits to a three-level model (solid lines) yield $g^{(2)}(0) \leq 0.5$ in both cases, demonstrating the retention of single-photon purity and emitter functionality after exposure to elevated temperatures.
\textbf{d-f}, Optical images of exfoliated hexagonal boron nitride (hBN) flakes transferred on a Si/SiO$-2$ substrate, labeled as Flake 1, Flake 2, and Flake 3, with approximate thicknesses of 63 nm, 150 nm, and 421 nm, respectively.
\textbf{g-i}, Corresponding second-order autocorrelation measurements,$g^{(2)}(\tau )$, acquired from representative emitters located at irradiation spots with a dose of 40 s in each flake. The data are fitted using a three-level system model, yielding $g^{(2)}(0)$ values of $0.16 \pm 0.06$, $0.12 \pm 0.04$, and $0.09 \pm 0.06$ for Flake 1, Flake 2, and Flake 3, respectively, confirming single-photon emission behavior.
}
\label{fig:figure3}
\end{figure*}
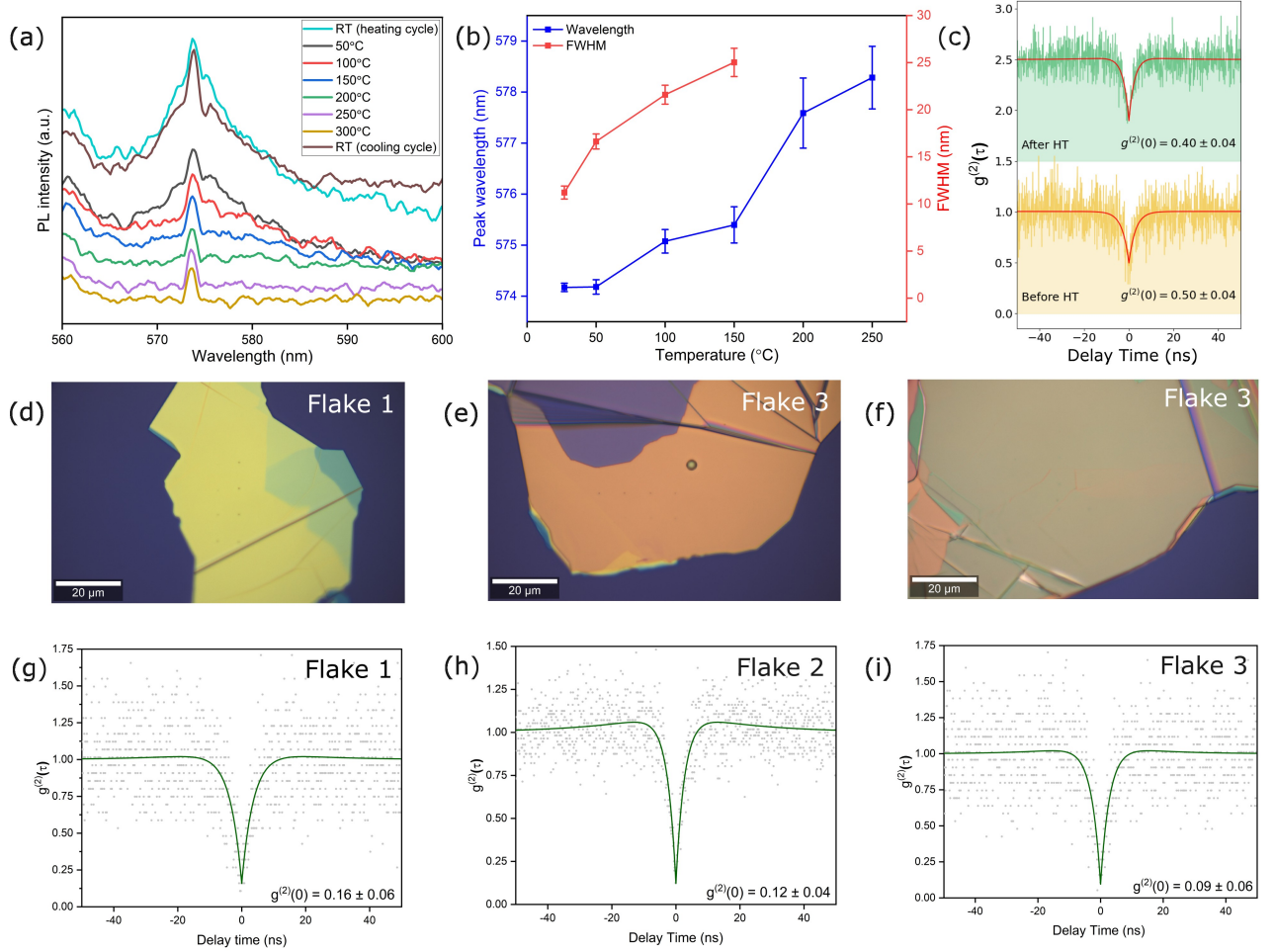

\begin{figure*}[t]
\centering

\begin{picture}(0,0)
\put(-5,80){\textbf{}}
\end{picture}
\includegraphics[width=0.95\textwidth]{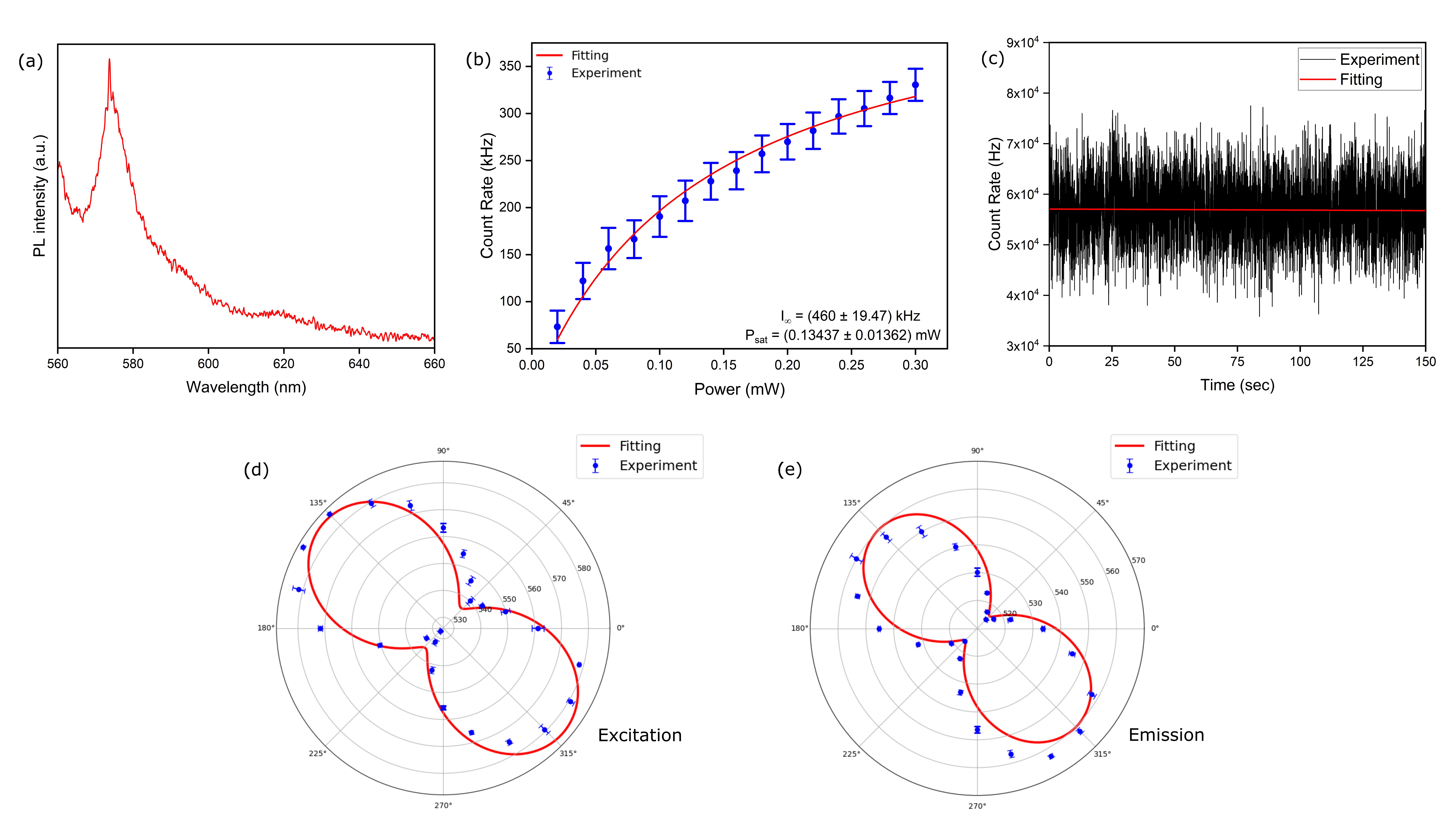}\hspace{0.015\textwidth}

\caption{\textbf{Optical characterization of a single-photon emitter (SPE) in hBN}
\textbf{a}, PL spectrum of an individual SPE, recorded using 532 nm excitation. The emitter exhibits an emission line centered at approximately $\sim$574 nm.
\textbf{b}, Excitation power-dependent PL intensity of the emitter. The saturation behavior is well described by a two-level system model, yielding a saturation count rate of $I_{\infty} = (460 \pm 19.47) \times 10^3$ cps and a saturation power of $P_{\mathrm{sat}} = (134.37 \pm 13.62)$ $\mu W$.
\textbf{c}, Temporal stability of the emitter emission recorded at an excitation power of 100 $\mu W$.
\textbf{d-e}, Polarization-dependent excitation and emission measurements of the SPE, respectively. Blue symbols represent the experimental data, while the solid curves correspond to fits using the function $I(\theta)=a+b\sin^{2}(\phi + \theta_{0})$, indicating the dipole-like nature of the optical transitions.
}
\label{fig:figure4}
\end{figure*}

First part of Fig. 3 presents the thermal response of the fabricated single-photon emitters (SPEs) through in-situ photoluminescence (PL) measurements and second-order correlation analysis. In Fig. 3(a), the PL spectra are recorded as the sample temperature is systematically increased from room temperature (RT) to $300^\circ\mathrm{C}$ in increments of $50^\circ\mathrm{C}$ using a 50× objective lens (NA = 0.5). A clear temperature-dependent reduction in emission intensity is observed. The gradual decrease in PL intensity with increasing temperature is consistent with thermally activated nonradiative recombination pathways, which compete with radiative emission and lead to quenching of the optical signal. The emission remains well-defined up to approximately $200^\circ\mathrm{C}$, indicating good thermal robustness of the defect states within this regime. Beyond this temperature, the intensity is significantly suppressed, suggesting enhanced activation of nonradiative channels or phonon-assisted processes. To further elucidate the thermal stability of the emitter, the evolution of the PSB peak wavelength and full width at half maximum (FWHM) extracted from the spectra is summarized in Fig. 3(b). As the temperature increases from room temperature to $250^\circ\mathrm{C}$, the PSB peak exhibits a gradual red shift from approximately 574 nm to 578 nm. Such a temperature-induced shift is commonly attributed to modifications of the local defect environment, including thermally induced lattice expansion, changes in local strain, or variations in electron–phonon coupling. Notably, the total spectral displacement remains relatively small $(<5) nm$ over the investigated temperature range, indicating that the electronic structure of the defect remains largely preserved during thermal cycling.

From the linewidth evolution plot, the FWHM initially increases with temperature, reaching a maximum value of approximately 25 nm around $150^\circ\mathrm{C}$. This broadening is consistent with enhanced electron–phonon interactions and thermally activated dephasing processes that become increasingly significant at elevated temperatures. The emission intensity decreases substantially at temperatures above $150^\circ\mathrm{C}$, leading to a reduced signal-to-noise ratio and larger uncertainties in the spectral fitting procedure. Consequently, the FWHM values obtained at the highest temperatures, particularly at $200^\circ\mathrm{C}$ and above, should be interpreted with caution, as the weak emission signal limits the reliability of the linewidth extraction.

Importantly, this thermally induced quenching is found to be reversible. Upon cooling the sample back to room temperature (denoted as RT(cooling cycle) in Fig. 3(a)), the emission intensity and spectral profile are largely restored, demonstrating that no permanent structural damage occurs within the investigated temperature range. This reversible behavior is fundamentally distinct from irreversible degradation mechanisms reported at higher annealing temperatures $(>500^\circ\mathrm{C})$ \cite{hazra2026insights}, where permanent loss of emission and modification of defect lifetimes have been observed. The present results, therefore, indicate that moderate thermal cycling does not compromise the intrinsic optical properties of the emitters.

Fig. 3(c) further validate the preservation of single-photon emission characteristics through second-order autocorrelation measurements, $g^{(2)}(\tau )$, acquired at room temperature before and after the thermal cycle, respectively. Both measurements exhibit clear antibunching behavior with $g^{(2)}(0) \leq 0.5$, confirming the non-classical nature of the emission. The three-level model fits (solid curves) show good agreement with the experimental data, indicating that the emitter dynamics remain governed by the same underlying photophysical processes before and after heating. The small change in $g^{(2)}(0)$ values demonstrates that the single-photon purity is preserved, and no additional background emission or multi-photon contributions are introduced by the thermal treatment. The observed thermal stability and reversibility are particularly significant for practical applications. In environments such as space-based quantum communication systems, components are subjected to wide and cyclic temperature variations, ranging from cryogenic conditions in shadowed regions to elevated temperatures under solar exposure\cite{yin2017satellite, liao2017satellite}. The ability of these SPEs to retain their optical functionality and single-photon purity after thermal cycling up to $300^\circ\mathrm{C}$ highlights their resilience and suitability for such demanding conditions. The maintenance of emission characteristics up to at least $150 - 200^\circ\mathrm{C}$ is especially relevant for ensuring reliable operation in satellite-based quantum key distribution (QKD) and related photonic technologies.

In the second part of Fig. 3, Second-order autocorrelation measurements, $g^{(2)}(\tau )$, were performed on hBN flakes of varying thicknesses to evaluate the single-photon purity as a function of flake thickness. Optical images of three representative flakes are shown in Fig. 3(d–f). The flake thickness was measured using a surface profilometer by evaluating the step height between the flake and the substrate. The height profiles (see in the supplementary material, Sec. S1) reveal substantial thickness variations, with Flake 1, Flake 2, and Flake 3 measuring 63 nm, 150 nm, and 421 nm, respectively. The second-order autocorrelation measurements, $g^{(2)}(\tau )$, for the three flakes are presented in Fig. 3(g–i). All three sites exhibit pronounced antibunching at zero delay, with fitted using a three-level system model, yielding $g^{(2)}(0)$ values of $0.16 \pm 0.06$, $0.12 \pm 0.04$, and $0.09 \pm 0.06$ for Flakes 1, 2, and 3, respectively. Clearly fall below the 
$g^{(2)}(0)$ < 0.5 threshold, confirming the presence of high-purity single-photon emitters (SPEs). A primary finding of this work is the robust formation of high-purity emitters across a wide range of flake thicknesses. While it has been suggested that hBN emitter performance is independent of the host thickness\cite{kumar2023localized}, our findings further substantiate this stability within significantly thicker flakes. This observation opens new avenues for exploring the scalability of hBN-based emitters in practical applications.

This thickness independence is essential for the fabrication of monolithic quantum photonic integrated circuits (QPICs)\cite{o2009photonic, wang2020integrated}. Furthermore, thicker flakes offer the superior structural integrity essential for implementing advanced nanofabrication techniques, such as lithographic patterning, waveguide integration, and cavity coupling\cite{spencer2023monolithic, li2021integration}. A scalable and varied approach to integrating quantum light sources into reliable, high-performance integrated circuits is made possible by this stability across several dimensions.

Fig. 4a presents the PL spectrum of a color center emitter, characterized by a pronounced emission centered at 574.3 nm and a linewidth of 14.1 nm (FWHM). Recent spectral reassignment reveals that the prominent room-temperature emission near $\sim$575 nm, previously attributed to the zero-phonon line (ZPL), is actually the phonon sideband (PSB)\cite{hazra2026temperature} of the color center defect. The true ZPL lies near $\sim$548 nm but becomes significantly broadened at ambient conditions, causing the PSB to dominate the observed spectrum. An additional consideration arises from the excitation conditions. Under 532 nm laser excitation, the intrinsic Raman mode of hBN at 1366 cm$^{-1}$ corresponds to an emission wavelength of $\sim $574 nm, which spectrally overlaps with the observed peak. Given that this Raman signal lies within the detection window (e.g., a 550 nm long-pass filter), it may contribute to or partially overlap with the measured emission. Therefore, the presence of a $\sim$ 14.1 nm-wide feature at this wavelength necessitates careful discrimination between defect-related emission (PSB) and Raman scattering. To unambiguously confirm the origin of the peak, further validation is required. In particular, excitation-wavelength-dependent measurements would reveal a shift in the Raman peak while leaving defect emission unchanged. Alternatively, power-dependent studies can provide a clear distinction: Raman scattering intensity scales linearly with excitation power, whereas defect emission typically exhibits saturation behavior. This distinction is supported by the excitation power–dependent measurements shown in Figure 4b. The emission count rate increases nonlinearly with increasing excitation power and follows a clear saturation trend. The data were fitted using the relation $ I = I_{\infty} P / (P + P_{\text{sat}}) $, where $ I_{\infty} $ represents the saturation emission rate and $ P_{\text{sat}} $ denotes the saturation power\cite{sakib2024purcell}. From the fit, a maximum emission rate of $ (460\pm 19.47) \times 10^{3} $ counts/s was obtained at a saturation power of  $(134.37\pm 13.62)\mu W$. The observed sublinear increase at higher excitation powers confirms that the emission originates from a localized quantum system with a finite excited-state lifetime, rather than from Raman scattering, which would exhibit a strictly linear dependence.

Beyond spectral purity and brightness, key performance indicators for single-photon emitters (SPEs) include photostability and polarization properties. The photostability of the emitter was evaluated by monitoring its emission intensity over a duration exceeding 150 s at an excitation power of 100 $\mu W$. As shown in Figure 4c, the emission remains stable without any observable blinking or photobleaching, confirming the excellent photostability of the emitter.

To probe the polarization characteristics, an analyzer was introduced in the detection path while maintaining a fixed excitation polarization. The results indicate that the emission is linearly polarized. To further assess whether the emission originates from a single dipole or multiple dipoles, both excitation and emission polarization measurements were performed. The polarization data (Fig. 4d and 4e) were fitted using the function $ a + b \sin^{2}(\phi + \theta_{0}) $, where $\theta_{0} $ is the polarization angle. The extracted angles for excitation $138.6 \degree$ and emission $135\degree$ are nearly identical, with corresponding polarization visibilities of $4.39 \% $ and $4.08 \%$, respectively. However, it must be noted that such notably low polarization visibilities (approximately 4$\%$) offer relatively limited evidence to conclusively claim a single, linearly polarized dipole. It is quite possible that the observed modulation depth is artificially decreased by increased background signals, which could be derived from carbon atoms. As a result, the near-equivalence of the excitation and emission polarization angles provides a far more reliable and convincing indicator of an underlying single-dipole-like transition, even though the raw visibility data should be interpreted with caution.

\section{Conclusion}
We demonstrate that focused electron-beam irradiation enables the reproducible and deterministic generation of room-temperature single-photon emitters in hBN, exhibiting high single-photon purity with $g^{(2)}(0)=0.09$--$0.16$ across three distinct flakes. Notably, the consistent formation of high-purity emitters across flakes spanning a wide thickness range highlights the robustness and scalability of this approach. Based on the revised spectral assignment, the prominent emission feature near $\sim$ \SI{575}{nm} at room temperature corresponds to the phonon sideband of a green–yellow emitter, with the true zero-phonon line located around $\sim$ \SI{548}{nm}. The key advances of this work are twofold: first, the establishment of a quantitative correlation between electron irradiation dose and single-emitter performance; and second, the observation of reversible thermal quenching behavior up to \SI{300}{\degreeCelsius}. This reversible thermal response is fundamentally different from the irreversible degradation induced by high-temperature annealing reported previously \cite{hazra2026insights}. Collectively, these findings expand the deterministic fabrication strategies for hBN-based quantum light sources by introducing precise dose control and demonstrating operational stability beyond room temperature.

\section{Methods}
\subsection{hBN flakes transfer and thickness measurement}
A standard Si/SiO$_2$ substrate with a thermally produced oxide layer of $300$ nm was used. The substrates that were prepared for electron beam irradiation underwent photolithography procedures and metal deposition to create a hard mask grid. The grid made it easy to find the hBN flake for electron beam irradiation. First, the substrate was cleaned by sonicating it in acetone, IPA, and DI water for two minutes. After drying it with N$_2$ blow, it was heated to $180 \degree C$ on a hot plate for 10 minutes. Nitto blue tape was used to transfer exfoliated hexagonal boron nitride (hBN) flakes onto a handmade polydimethylsiloxane (PDMS) film, which was formed by combining PDMS solution with curing agent in a 10:1 ratio and letting the mixture cure overnight. After that, these hBN flakes were then placed onto a grid-marked Si/SiO$_2$ substrate. A DektakXT profilometer was used to measure the hBN flakes' thickness. A precise measurement of the height difference between the two surfaces was obtained by scanning the instrument's $2 \mu m$ diameter tip from a bare substrate region to the area holding the flake. The thickness was measured along the dotted line in the optical images in the supplementary material, Sec S1.

\subsection{Electron beam irradiation}
Defect centers were created by exposing hexagonal boron nitride (hBN) flakes to an electron beam via a Scanning Electron Microscope (SEM) system. The sample was positioned at a working distance of $4.5,\mathrm{mm}$, and all experiments were performed with an electron beam current of $48.2,\mathrm{pA}$ and an acceleration voltage of $10,\mathrm{kV}$. Each experiment was carried out in a high-vacuum environment with a pressure of $9.6 \times 10^{-6},\mathrm{Pa}$. The position selection is a manual method that involves adjusting the x and y values to find the desired place, which may result in regular spacing. The electron dose (irradiation time), which varied from 20 to 180 seconds, was utilised to adjust the electron dose. An electron dose of up to 100 seconds was discovered to be optimal for efficient emitter creation. Directing the electron beam to a single point resulted in localised emitter production, as demonstrated in the PL map in Fig. 2a.

\subsection{Optical characterization}
A Witec alpha 300RA modular confocal Raman microscopy system was equipped with a CW 532 nm laser for both excitation and scanning. A large numerical aperture (NA = 0.9, Zeiss) objective lens was used to focus the laser onto the sample after it had passed through a half-waveplate. The silicon reference peak at $520.7,\mathrm{cm}^{-1}$ was used to calibrate the Raman spectrometer. An X-Y-Z closed-loop stage was used for scanning. A 532 nm dichroic mirror and an additional 550 nm long pass filter (Thorlabs) were used to filter the collected emission. A flipping mirror in the emission channel directs the signal to two single photon avalanche diodes (SPADs) (Excelitas Technologies) for single photon counting or to a spectrometer (UHTS 300 SMFC Green-NIR, WiTec focus innovations). Correlation measurements were performed using a time-correlated single photon counting module (Time Tagger Ultra, Swabian instruments), with the sample stimulated by a 532nm CW laser and the signal recorded using SPADs (dark current = 100 counts/s). Background correction was not taken into account while showing the $g^{(2)}(\tau )$ curves.

Fluorescence Lifetime Imaging Measurement (FLIM) was done on a MicroTime 200 (PicoQuant, GmbH, Germany) applying the time-tagged time-resolved (TTTR) methodology, allowing for the reconstruction of PL decay traces from emitter spots in the confocal volume. The arrangement is connected to an inverted microscope (Olympus IX71) with a water immersion objective (UPlanSApo NA 1.2, 60×, WD = 0.28 mm). A pulsed diode laser (PDL 828 Sepia II, PicoQuant) with a full width at half-maximum of 176 ps and a repetition rate of 20 MHz has been used to trigger emitters at 532 nm. The collected fluorescence is focused onto a $50 \mu m$ pin-hole after being filtered using a 580 band-pass filter (AHF/Chroma, Germany) to exclude the excitation light. A single photon avalanche photodiode (SPAD) received the collected fluorescence signal. The sample was raster-scanned in order to image a region. Typically, a resolution of 512 × 512 pixels and a collection duration of 0.70 ms per pixel were used to obtain the image data.

High-temperature PL spectroscopy was performed from ambient temperature to $300^\circ\mathrm{C}$, making use of a closed-cycle refrigerating system cryostat (Janis CCS-XG-M/204N) in a Laser Raman Spectroscopy system (LRS: HR800-UV confocal micro-Raman spectrometer). Photoluminescence measurements were carried out applying a 532 nm excitation and a 50x microscope objective lens with a numerical aperture of 0.5.

\subsection{AFM measurement}
Atomic force microscopy (AFM) measurements were performed using an Oxford Instruments Asylum Research MFP-3D BIO system (Oxford Instruments Asylum Research Inc., USA). Surface morphology was characterized in tapping mode (AC Air mode) under ambient conditions. Images were acquired with a scan rate of 0.5 Hz and a resolution of 512 points per scan line. The proportional and integral feedback gains were maintained in the range of 0.3-0.6 to ensure stable imaging. The cantilever drive frequency was typically between 300 and 335 kHz, with a drive amplitude of approximately 24 mV and a set-point voltage of 0.5-1.0 V. The acquired AFM images were processed using second-order flattening to remove background tilt and scanner-induced artefacts. The root-mean-square (RMS) surface roughness (Rq) was subsequently calculated using Asylum Research software.

\section*{ acknowledgments}
We acknowledge funding support from the National Quantum Mission, an initiative of the Department of Science and Technology, Government of India. We also acknowledge funding from ANRF via grant number SPR/2023/000175. A.M., J.K. and I.S. acknowledge institute teaching assistantship from IIT Bombay, India. R.K. acknowledges University Grants Commission for research fellowship. We acknowledges the Centre for Sophisticated Instruments and Facilities (CSIF) and the Centre of Excellence in Nanoelectronics (CEN), IIT Bombay for providing access to fabrication and sample characterization.

\section*{ Author Contributions}
\textbf{Amrita Majumder:} Prepared the samples; Performed the irradiation; quantum optical characterization; Data analysis; and writing - original draft. \textbf{Janhavi Khunte:} PL/Raman characterizations; Data analysis; Prepared markers on substrate. \textbf{Ikshvaku Shyam:} AFM measurements. \textbf{Rohit Kumar:} SEM imaging. \textbf{Anshuman Kumar:} Funding acquisition; Project supervision.

\bibliography{sources.bib}
\end{document}